# The role of ionized impurity scattering on the thermoelectric performances of rock salt AgPb$_m$SnSe$_{2+m}$

*Lin Pan$^{1, 4*}$, Sunanda Mitra$^2$, Li-Dong Zhao$^3$, Yawei Shen$^1$, Yifeng Wang$^{1*}$ Claudia Felser$^4$ and David Berardan$^{2*}$*

L. Pan, Y. Shen, Y. Wang
College of Materials Science and Engineering, Nanjing Tech University, China, 210009,
E-mail: linpan@njtech.edu.cn, yifeng.wang@njtech.edu.cn
S. Mitra, D. Berardan
SP2M – ICMMO (UMR CNRS 8182), Université Paris-Saclay, Univ. Paris-Sud, F-91405 Orsay, France
E-mail: david.berardan@u-psud.fr
L-D. Zhao
School of Materials Science and Engineering, Beihang University, Beijing 100191, China
C. Felser
Max Planck Institute for Chemical Physics of Solids, D-01187 Dresden, Germany

We report on the successful synthesis and on the properties of polycrystalline AgPb$_m$SnSe$_{2+m}$ (m = ∞, 100, 50, 25) samples with a rock salt structure. Between ~ 160 K and ~ 400 K, the dominant scattering process of the carriers in this system changes from acoustic phonon scattering in PbSe to ionized impurity scattering in AgPb$_m$SnSe$_{2+m}$, which synergistically optimizes electrical and thermal transport properties. Thanks to the faint amount of AgSnSe$_2$, the Seebeck coefficient is enhanced by boosting the scattering factor, the electric conductivity is improved by the increase of the concentration of holes coupled to a limited degradation of their mobility, and the total thermal conductivity is reduced by suppressing bipolar thermal conductivity. Therefore, ZT of AgPb$_m$SnSe$_{2+m}$ (m = 50) reaches 1.3 at 889 K. The mechanism suggested in this study opens new paths to improve the thermoelectric performances of other families of materials.

## 1. Introduction

Thermoelectric (TE) effect enables direct conversion between thermal and electrical energy and provides an alternative route for power generation and refrigeration[1]. The TE conversion efficiency of TE materials is an increasing function of the dimensionless figure of merit $ZT = (S^2T) / (\rho\lambda)$, where $S$, $\rho$, $\lambda$ and $T$ are the Seebeck coefficient, the electrical resistivity, the thermal conductivity and the absolute temperature, respectively[2]. Therefore, excellent thermoelectric materials require a perfect combination of high $S$ with low $\rho$ and $\lambda$. However, the Seebeck coefficient is related to the electrical conductivity according to the Boltzmann transport equations[3], hence, it will be hard to optimize one without degrading the other. Many approaches have been developed in the recent years to enhance the thermoelectric power factor ($S^2/\rho$) including modifying the band structure by electronic resonant states[4], quantum confinement effects[5], band convergence[6], energy barrier filtering[7, 8] and intensifying impurity scattering of the carriers[9] to enhance the Seebeck coefficient. Among these, the carrier impurity scattering effect on the Seebeck coefficient has been recently investigated in Skutterudite[9] after it had nearly been ignored during the past half century. The carrier scattering probability by ionized impurity is inversely proportional to $\varepsilon^{3/2}$ and $v^3$, where $\varepsilon$ is the energy of the carriers and $v$ is the drift velocity[10]. Therefore the low energy carriers are more scattered than the high energy ones leading to a positive energy dependence of the scattering time, which can in turn enhance the Seebeck coefficient. On the other hand, it is worthy to point out that strong ionized impurity scattering will significantly reduce the carrier mobility which thus leads to a deterioration of the electrical conductivity[11]. Hence, the key point to use carrier ionized impurity scattering for improving the thermoelectric performances is to find an effective strong ionized impurity scattering center, which could simultaneously enable to tune the concentration of carriers in order to reduce the degradation of the electrical conductivity due to much lower mobility. Typically, the lattice of the ionized impurity scattering center should match well with the matrix, which could minimize the detrimental effects on electrical transport properties.

Recently, AgSnSe$_2$ with rock salt crystal structure, which is identical to PbSe, has attracted much attention as a natural valence-skipping (the Sn ions separated into 1:1 mixture of Sn$^{2+}$ and Sn$^{4+}$) three-dimensional superconductor with quantum phase fluctuations[12, 13]. In the normal state, the electrical resistivity of AgSnSe$_2$ is around 0.23 mΩ·cm at room temperature which originate from high carrier concentration of $2.0 \times 10^{22}$ cm$^{-3}$. Besides, the mean-free path of the carriers in AgSnSe$_2$ with strong disorder is around 0.9 nm, which is very short, implying the existence of intrinsically strong electron scattering possibly due to the valence fluctuations of Sn[13]. On the other hand, rocksalt-structured PbSe was considered as an alternative for its analogue PbTe due to the lower cost of Se as compared to Te, and higher melting and operation temperature[14]. Nevertheless, compared with PbTe, the electronic band structure of PbSe is not favorable for TE performance since the two key valence bands are farther apart in PbSe than in PbTe. The light-hole L to heavy-hole Σ band separation in PbSe is 0.26 eV compared to 0.15 eV in PbTe[15]. Although solid solution had been verified as a very promising way to realize band convergence to get high Seebeck coefficient and meanwhile decrease the thermal conductivity through point defect engineering[6, 7, 16], there is no clear evidence that ionized impurity scattering can optimize thermoelectric performance in solid solution. Herein, in this paper, we report the thermoelectric properties of a novel PbSe-AgSnSe$_2$ solid solution AgPb$_m$SnSe$_{2+m}$, with promising performances driven by ionized impurity scattering of the charge carriers.

## 2. Results and discussion

X-ray diffraction patterns at room-temperature for all the samples are shown in **Figure 1** (a). It can be seen that for AgPb$_m$SnSe$_{2+m}$, the main diffraction peaks, which are very sharp correspond to PbSe (JCPD No.06-0354) phase, no other phase being detected obviously from the XRD patterns. Accurate lattice parameters calculated from the XRD patterns using Rietveld refinement, shown in Figure 1(b), indicate that the lattice constant *a* monotonically decreases with increasing AgSnSe$_2$ content (see Figure S1 of Supporting Information for Rietveld refinements of these XRD patterns). Low-magnification scanning electron microscopy images of the fracture surfaces of typical samples (Figure S2) coupled to energy dispersive spectrometer (EDS) analysis show that the obtained samples have low porosity, and that the actual composition is in good agreement with the nominal composition. These results show that AgSnSe$_2$ has been successfully introduced into PbSe, leading to the formation of AgPb$_m$SnSe$_{2+m}$.

The temperature dependences of (a) electrical resistivity $\rho$ and (b) Seebeck coefficient *S* for AgPb$_m$SnSe$_{2+m}$ (m = ∞, 100, 50, 25) in the range of 25 K − 889 K are shown in **Figure 2**. A good agreement can be noticed between the low and high temperature measurement results using two different measurement setups. Increasing $\rho$ and *S* with rising temperature from 50 K to 595 K and the positive values of *S* indicate a *p*-type degenerate semiconductor behavior below 595 K for pristine PbSe. However, *S* reaches a maximum and starts to decrease when heating above 595 K and even changes to negative values around 840 K. Meanwhile, $\rho$ begins to drop when rising temperature to 693 K. This electrical transport behavior unambiguously corresponds to a two bands conduction mechanism, with both holes and electrons due to thermal activation of the carriers through the band-gap, which fully agrees with previous report[17] (as holes and electrons have opposite contributions to the Seebeck coefficient, bipolar conduction with the simultaneous presence of both type of carriers is detrimental for thermoelectric performances). Indeed, the thermal activation of carriers, which could also contribute to an evolution of the temperature dependence of S, could not turn the positive values observed at low temperature into negative at high temperature. Contrary to PbSe, $\rho$ of AgPb$_m$SnSe$_{2+m}$ (m = 100, 50, 25) first increases when heating and then decreases, leaving a flat peak around 173 K, 187 K and 201 K respectively. At a first glance, this non-monotonous temperature dependence of $\rho$ could correspond to a thermally activated semiconductor behavior. In that framework, the activation energy $\Delta E$ for conduction of AgPb$_m$SnSe$_{2+m}$ (m = 100, 50, 25) in the intermediate temperature range would be 0.028 eV, 0.052 eV, and 0.054 eV respectively, obtained using: $\rho = \rho_0 \exp\left(\dfrac{\Delta E}{k_B T}\right)$, where $\rho_0$ is the pre-exponential factor, $k_B$ is Boltzmann's constant (J/K). With further increasing temperature, $\rho$ keeps on falling and then rises again, leaving a flat valley around 345 K, 401 K, and 399 K respectively, which would thus correspond to the saturation of the activation of the impurity band. Although $\rho$ of AgPb$_m$SnSe$_{2+m}$ is larger than the one of pristine PbSe at low temperature, it is noteworthy that it has been improved at high temperature due to this non-monotonous behavior. In addition, Enhanced $S$ without peak with rising temperature over the whole temperature range indicates that bipolar carriers conduction has been suppressed in AgPb$_m$SnSe$_{2+m}$ (m = 100, 50, 25) as compared to PbSe.

In order to get a better understanding of the role of AgSnSe$_2$ in the system, Hall-effect analysis was performed. Assuming one carrier type (which is also valid for pristine PbSe below 600-700K) and a simple parabolic band model in our analysis[18], carriers concentration [$n$] for AgPb$_m$SnSe$_{2+m}$ (m = $\infty$, 100, 50, 25) has been calculated using the measured Hall coefficients, $R_H$, following the relationship $R_H = 1 / ne$, where $e$ is the electronic charge, as shown in Figure 2(c) (although the values obtained with these hypothesis are estimations, it enables good direct comparison between the different samples and the analysis of the temperature dependences). [$n$] of all samples decreases with rising temperature, which is similar to previously work on $p$ type PbSe. It reveals that interband excitations from heavy-hole band $\Sigma$ to light-hole band L of AgPb$_m$SnSe$_{2+m}$ (m = $\infty$, 100, 50, 25) are absent or not obvious below room temperature[19]. With the increase of AgSnSe$_2$ content, the carrier concentration of AgPb$_m$SnSe$_{2+m}$ increases monotonically over the entire temperature range, from 0.22×10$^{19}$ cm$^{-3}$ for pristine PbSe to 2.58×10$^{19}$ cm$^{-3}$ for AgPb$_m$SnSe$_{2+m}$ (m = 25) at room temperature as shown in **Table 1**. Our pristine PbSe sample is slightly off-stoichiometric, due to slight excess of Se during the synthesis process, which leads to the metallic behavior (or degenerate semiconductor). When increasing the AgSnSe$_2$ content, more Se will be available, and the PbSe phase might be a little bit more Se-enriched, resulting in a larger concentration of carriers, a lower *S*, and still a metallic behavior at low temperature. However, the most striking observation that can be made in Figure 2(c) is the monotonous temperature dependence of the concentration of carriers of AgPb$_m$SnSe$_{2+m}$ (m = 100, 50, 25), which rules out the "thermal activation" behavior we suggested above to explain the temperature dependence of the electrical conductivity. Indeed, if the presence of AgSnSe$_2$ led to an impurity band a few meV above the valence band maximum, leading to thermal activation in the temperature range 170-200K, a distinct anomaly should be observed in the [*n*](T) curve, which is not the case here. Besides, this former explanation would not be consistent with the much larger value of the residual resistivity of AgPb$_m$SnSe$_{2+m}$ (m = 100, 50, 25) observed at low temperature, although the carriers concentration is much larger than in the pristine PbSe sample.

Resistivity $\rho$ is given by $\rho = \dfrac{1}{ne\mu}$, where *n* is the carriers concentration, *e* is the electronic charge (C), and $\mu$ is the mobility of the holes. The temperature dependence of the mobility of all the samples is shown in the Figure 2(d) which scales with the power exponent $\delta$ of $\mu \approx T^{-\delta}$. $\mu$ decreases significantly over the entire temperature range with the addition of AgSnSe$_2$, from 2690 cm$^2$V$^{-1}$s$^{-1}$ for pristine PbSe to 40 cm$^2$V$^{-1}$s$^{-1}$ for AgPb$_m$SnSe$_{2+m}$ (m = 25) at room temperature, as shown in Table 1, which is obviously at the origin of the larger low temperature electrical resistivity observed with (m = 100, 50, 25). The mobility $\mu$ of pristine

PbSe decreases monotonically over the entire temperature range, while $\mu$ of AgPb$_m$SnSe$_{2+m}$ (m = 100, 50, 25) decreases first and then increases, leaving a flat valley around 160 K (m = 100) and 180K (m = 50, 25) respectively. It is worthy to point out that with further heating, $\mu$ of AgPb$_m$SnSe$_{2+m}$ (m = 100) keeps climbing first and then falls, leaving a maximum around 350 K. Actually, the evolution of $\mu$ of AgPb$_m$SnSe$_{2+m}$ (m=50, 25) is identical to that of AgPb$_m$SnSe$_{2+m}$ (m = 100). In other words, $\mu$ of AgPb$_m$SnSe$_{2+m}$ (m = 50, 25) should also exhibit a peak located around 400 ~ 450 K. These two peaks cannot be observed in Figure 2(d) due to the temperature limit of the measurements. However, they can be deduced from the temperature dependence of the lattice thermal conductivity, as discussed later.

The mobility can be expressed as a function of a relaxation time $\tau$ and a transport effective mass $m^*$: $\mu = e\tau / m^*$. Here, we assume $m^*$ is an invariant (relative to temperature or energy), which of course constitutes a rough approximation but is sufficient for the discussion. Regarding the relaxation time $\tau$, it can be decomposed following Matthiesen's rule as:

$$\tau^{-1} = \tau_{h-h}^{-1} + \tau_{h-GB}^{-1} + \tau_{h-II}^{-1} + \tau_{h-ph}^{-1} + ...$$

(where $\tau_{h-h}$ refers to hole-hole scattering, h-GB to hole-grain boundaries scattering, h-II to hole-ionized impurities scattering, and h-ph to hole-phonons scattering). All these scattering process depend on the temperature and the concentration of carriers. $\tau_{h-h}$ can be neglected in our case as it is only predominant at very low temperatures and as the concentration of carriers in our system is rather low, $\tau_{h-ph}$ is almost always the main contributor to $\tau$ at high temperature, when most phonon modes are activated, and $\tau_{h-II}$ can be neglected in PbSe as the concentration of carriers is low (and so should be the concentration of ionized impurities). Under this hypothesis, $\mu$ follows a power law $\mu \approx T^{-\delta}$, where $\delta$=s-r is composed of the scattering parameter (i.e., $r = -1/2$ for acoustic phonon scattering, 0 for neutral impurity scattering, and 3/2 for ionized impurity scattering, …) and the number of phonons participating to a scattering process (i.e., $s = 1$ for one-phonon process, 2 for two-phonon process, and s = 0 for non-phonon scattering processes) [20]. Here, the power exponents $\delta$ close to room temperature for AgPb$_m$SnSe$_{2+m}$ (m = ∞, 100, 50, 25) is 1.8, -1.4, -1.5 and -1.7, respectively. More precisely, in the degenerate *p*-type PbSe sample, the mobility follows $\mu \approx T^{-1.2}$ at 80 – 160 K and $\mu \approx T^{-1.8}$ above 180 K, indicating $r = -1/2$ (acoustic phonon scattering with one-phonon process) over the entire temperature range (the slightly lower exponent observed at low temperature most probably originates from a contribution of h-GB scattering). For the degenerate AgPb$_m$SnSe$_{2+m}$ (m = 100, 50, 25) samples below 160 K, the mobility showed $\mu \approx T^{-0.35}, T^{-0.45}, T^{-0.5}$ indicating that the carrier scattering mechanisms are a combination of various mechanisms. However, with temperature rising above 180 K, the mobility of AgPb$_m$SnSe$_{2+m}$ (m = 100, 50, 25) samples follows $\mu \approx T^{-1.4}, T^{-1.5}, T^{-1.7}$ indicating that ionized impurity scattering becomes the main carrier scattering mechanism in these samples above 180 K. When further heating to 350 K, $\delta$ value of AgPb$_m$SnSe$_{2+m}$ (m = 100) changes to positive again (same behavior will occur around 400 ~ 450 K for AgPb$_m$SnSe$_{2+m}$ (m = 50, 25)), which means acoustic phonon scattering will become the dominated scattering again at high temperature, as most phonon modes get activated.

The strength of ionized impurity scattering is mainly determined by the concentration of impurity centers $N_i$, their charge states $\upsilon$, the carrier concentration $n$, the dielectric constant $\varepsilon$ and the density of states effective mass $m_d^*$ [21]. According to Brooks' and Herring's derivations, the carrier relaxation time of ionized impurity scattering is proportional to $m_d^{*1/2}$ and inversely proportional to $N_i$ and $\upsilon^2$. [9, 21] Thus a highly charged impurity with high screened Coulomb potential will more effectively scatter low energy electrons and lead to larger average electron energy and thus larger Seebeck coefficient. Here, strong ionized impurity scattering in the AgPb$_m$SnSe$_{2+m}$ system could originate from the valence fluctuations of Sn with charge states Sn$^{4+}$ and Sn$^{2+}$.

In a first approximation, the Seebeck coefficient of degenerate semiconductors is proportional to $m^*$ and inversely proportional to $n^{2/3}$ as:

$$S = \left(\frac{8\pi^{2/3} k_B^2 (r+3/2)}{3^{5/3} e h^2}\right)\left(\frac{m^*}{n^{2/3}}\right)T$$

Here, $h$ is Planck's constant (Js), $k_B$ is Boltzmann's constant (J / K), $T$ is the absolute temperature (K), and $n$ is the carrier concentration (cm$^{-3}$). Although this relation should be used when the concentration of carriers does not depend on the temperature, it can be used here to obtain a qualitative understanding of the transport behavior of the materials, as the temperature dependence of [n] is limited. From this relation, the effective mass $m^*$ of AgPb$_m$SnSe$_{2+m}$ (m = 100, 50, 25) and pristine PbSe, can be qualitatively estimated as shown in the Table 1. **Figure 3** shows the room temperature Pisarenko plot (Seebeck coefficient vs carrier concentration) for AgPb$_m$SnSe$_{2+m}$ (m = 100, 50, 25) and pristine PbSe. The solid black and red lines guide for eyes in this figure are based on a model employing a single parabolic band (SPB)[17, 22] with an effective mass $m^* = 1.3\ m_e$ with acoustic phonon scattering mechanism ($r = -1/2$), and $m^* = 1\ m_e$ with ionized impurity scattering mechanism ($r = 3/2$), respectively. Although these values are rough approximations, due to the SPB hypothesis, it shows that the introduction of AgSnSe$_2$ in PbSe increases the product $(r + 3/2) \cdot m^*$, due to a strong increase of $r$ coupled to an almost constant effective mass, which contributes to an enhancement of the Seebeck coefficient (at given concentration of carriers).

Using $\mu = e\tau / m^*$, the relaxation time $\tau$ at room temperature can be qualitatively estimated, as shown in Table 1. It can be seen that the relaxation time $\tau$ decreases dramatically at room temperature when a slight fraction of AgSnSe$_2$ is added to PbSe. Thus, we can now confidently explain the previous apparent discrepancy between carriers concentration and resistivity of AgPb$_m$SnSe$_{2+m}$ (m = 100, 50, 25) and pristine PbSe samples as well as the non-monotonous temperature dependence of the electrical resistivity. When AgSnSe$_2$ is introduced into the PbSe matrix, the main carriers scattering mechanism below room temperature changes, from acoustic phonon scattering for PbSe to ionized impurity scattering for AgPb$_m$SnSe$_{2+m}$ (m = 100, 50, 25). Ionized impurity scattering largely reduces the relaxation time $\tau$, resulting in a much lower mobility $\mu$, further leading to a larger resistivity, but also leads to the positive temperature dependence of $\mu$ below room temperature that partially limits the increase of the electrical resistivity.

The heat capacity $C_p$ of pristine PbSe and AgPb$_m$SnSe$_{2+m}$ (m = 25) as a function of temperature from 2 K to 400 K is presented in **Figure 4**. Obviously, $C_p$ of pristine PbSe and AgPb$_m$SnSe$_{2+m}$ (m = 25) are identical over the whole temperature range, (we have also checked the room temperature values of m = 50 and m = 100 and both are the same). The heat capacity of pristine PbSe reaches a value of 50 J/mol K at room temperature which is exactly the same as the expected classical high-T Dulong-Petit lattice heat capacity value $C_p \sim C_v = 3nR = 6R = 49.9$ J/mol K at constant volume V[23, 24], where R is the molar gas constant and n = 2 is the number of atoms per formula unit (f.u.). The inset in Figure 3(a) shows the low-T data of pristine PbSe and AgPb$_m$SnSe$_{2+m}$ (m = 25) plotted as $C_p / T$ versus $T^2$ allowing a conventional fit by[23]

$$\frac{C_p(T)}{T} = \gamma + \beta T^2 \quad (3)$$

Here, $\gamma$ is the Sommerfeld electronic linear specific heat coefficient and $\beta$ is the coefficient of the Debye $T^3$ lattice heat capacity at low temperature, which is obtained by a linear fit of the data below 3.5 K according to equation above. One can see that $\beta$ is nearly a constant, the variable range is between 1.02 mJmol$^{-1}$K$^{-4}$ and 0.92 mJmol$^{-1}$K$^{-4}$ which is of the order of the uncertainty of the measurement. $\gamma$ is close to zero for all samples (within the error bar of the measurement), which shows that the Fermi level is close to the band edge of the valence band, consistent with the moderate carrier concentration and a single-band conduction.

The value of the Debye temperature $\theta_D$ of pristine PbSe and AgPb$_m$SnSe$_{2+m}$ (m = 25) is obtained from $\beta$ using the relation[24] $\theta_D = (\frac{12\pi^4 N_A k_B n}{5\beta})^{1/3}$, where N$_A$ is Avogadro's number and k$_B$ is Boltzmann's constant, yielding $\theta_D$ of all the samples around 153 K. This result shows that introduction of AgSnSe$_2$ does not change the Debye temperature of PbSe. Obviously, the introduction of such small concentrations of AgSnSe$_2$ was not expected to influence strongly the phonon spectra of these materials at low temperature.

The temperature dependence of the thermal conductivity $\lambda$ for AgPb$_m$SnSe$_{2+m}$ (m = ∞, 100, 50, 25) is plotted in **Figure 5(a).** At room temperature, $\lambda$ of pristine PbSe is 1.82 Wm$^{-1}$K$^{-1}$. It decreases first due to the intensified phonon-phonon umklapp processes and then increases with heating, leaving a valley around 570 K. This behavior originates from bipolar diffusion i.e. more minority carriers (electrons here) jump across the band in the narrow band pristine PbSe with rising temperature and the formed diffusing electron-hole pairs lead to an additional thermal conductivity contribution (see later). However, $\lambda$ for AgPb$_m$SnSe$_{2+m}$ (m = 100, 50, 25) decreases with heating over the whole temperature range which means that the bipolar contribution to the thermal conductivity has been suppressed. This bipolar contribution can be suppressed by either increasing the majority carrier concentration through heavy doping or building up energy barrier filters. According to previous discussions, in the AgPb$_m$SnSe$_{2+m}$ system, the suppressed bipolar contribution may originate from both the enhanced majority carrier concentrations and increased scattering factor by ionized impurity. Comparing with pristine PbSe, the total thermal conductivity $\lambda$ of AgPb$_m$SnSe$_{2+m}$ (m = 50, 25) drops remarkably in the whole temperature range while $\lambda$ of AgPb$_m$SnSe$_{2+m}$ (m = 100) strengthened below 750K. For instance, $\lambda$ values of AgPb$_m$SnSe$_{2+m}$ (m = 100, 50, 25) and pristine PbSe are 1.22, 0.55, 0.68 and 1.32 WK$^{-1}$m$^{-1}$ at 874 K. The total thermal conductivity $\lambda$ is given by:[25] $\lambda = \lambda_{lat} + \lambda_c + \lambda_{bi} = \lambda_{lat} + L\sigma T + \lambda_{bi}$, where $\lambda_{lat}$, $\lambda_c$ and $\lambda_{bi}$ refer to lattice, carrier and bipolar contributions to the thermal conductivity (let us recall that $\lambda_{bi}$ is given by $\lambda_{bipolar} = \frac{\sigma_h \sigma_e}{\sigma_h + \sigma_e}(S_h - S_e)^2$), respectively, and $L$ is the Lorenz number which is obtained by fitting the Seebeck coefficient to the reduced chemical potential[2]. In order to clarify the contribution of $\lambda_{bi}$ at high temperature, the $\lambda_{bi}$ has been separated from the $\lambda$ according to the following method. The difference, $\lambda - L\sigma T$ as a function of $T^{-1}$ for the pristine PbSe is shown in Figure 5(b). Since the acoustic phonon scattering is predominant at low temperatures before bipolar diffusion becomes significant in lead chalcogenides[15], $\lambda_{lat}$ nearly equals to $\lambda - L\sigma T$, which is proportional to $T^{-1}$, approximately corresponding to: $\lambda_{lat} \propto \left(\frac{k_B}{h}\right)^3 \frac{a^4 \rho \theta_D^3}{\gamma^2 T}$, where $a$ is the lattice parameter, $\rho$ is the density, $\theta_D$ is the Debye temperature, and $\gamma$ is the Grüneisen parameter, which is a measure of the anharmonic nature of lattice vibrations[26].

As shown in Figure 5(b), $\lambda - L\sigma T$ of pristine PbSe decreases linearly when heating from room temperature, which confirms that acoustic phonon scattering is predominant. However, $\lambda - L\sigma T$ of $AgPb_mSnSe_{2+m}$ (m = 100) starts to decrease linearly above 350 K, while above 400 ~ 450K for $AgPb_mSnSe_{2+m}$ (m = 50, 25). It indicates that acoustic phonon scattering becomes the dominant scattering process above 350 K (m = 100) and 400 ~ 450K (m = 50, 25) for $AgPb_mSnSe_{2+m}$, which corresponds to the abnormal behavior of hole mobility. Besides, as the temperature is increased above 595 K, $\lambda - L\sigma T$ of pristine PbSe strongly deviates from the linear relationship between $\lambda_{lat}$ and $T^{-1}$ because the bipolar diffusion starts to contribute to the thermal conductivity. Comparing with pristine PbSe, $\lambda - L\sigma T$ of $AgPb_mSnSe_{2+m}$ (m = 100, 50, 25) starts to gradually deviate at 750-800 K for alloys.

The bipolar contribution $\lambda_{bi}$ at high temperatures was estimated by extrapolating the linear behavior, as shown in the inset of Figure 5(a). For example, $\lambda_{bi}$ can be clearly observed in pristine PbSe at high temperature, whereas it has been strongly suppressed over the whole temperature range for $AgPb_mSnSe_{2+m}$ (m = 100, 50, 25). This could partly explain why the total thermal conductivity $\lambda$ of $AgPb_mSnSe_{2+m}$ (m = 100, 50, 25) is lower than pristine PbSe, especially at high temperature. On the other hand, intensified mass fluctuation scattering from mass difference between Ag, Sn and Pb could also contribute to the reduction of $\lambda_{lat}$.

The figure of merit *ZT* resulting from the combination of the electrical and thermal transport properties over the range of 340 − 890K is shown in **Figure 6**. ZT of pristine PbSe reaches 0.49 at 344 K, climbs to a maximum of 0.6 at 448 K, and then starts to fall down to below 0.1 above 694 K due to bipolar contribution. Contrary to pristine PbSe, *ZT* monotonically increases with temperature in the whole temperature range for $AgPb_mSnSe_{2+m}$ samples, with a maximum *ZT* value at 889 K of 1.3 for $AgPb_mSnSe_{2+m}$ (m = 50). Therefore, our results indicate that the high temperature TE properties of PbSe can be remarkably improved by the introduction of slight amounts of $AgSnSe_2$ to form $AgPb_mSnSe_{2+m}$, by suppressing the bipolar contribution to the thermal conductivity while maintaining reasonably large *S* values due to ionized impurity scattering of the carriers. In addition, it is worthy to point out that $AgPb_mSnSe_{2+m}$ system possess good thermal stability, as shown in Figure S3.

## 3. Conclusion

In conclusion, we successfully synthesized $AgPb_mSnSe_{2+m}$ (m = 100, 50, 25) samples with a rock salt structure. The temperature dependence of the holes mobility *μ* combined with the difference between total and carrier thermal conductivity $\lambda - L\sigma T$, showed that ionized impurity scattering becomes predominant in $AgPb_mSnSe_{2+m}$ near room temperature (180 − 350 K for (m = 100), 180 − 400 ~ 450 K for (m = 50, 25)), although acoustic phonon scattering remains the main scattering mechanism at higher temperature. The introduction of $AgSnSe_2$ within the PbSe matrix can synergistically optimize the electrical and thermal transport properties. It enhances the Seebeck coefficient (at given carrier concentration) by boosting the scattering factor, improves the electric conductivity by the increase of the concentration of holes coupled to a limited degradation of their mobility, and reduces the total thermal conductivity by suppressing the bipolar contribution to the thermal conductivity.

Therefore, thanks to this introduction of ionized impurity scattering, ZT of $AgPb_mSnSe_{2+m}$ (m = 50) reaches 1.3 at 889 K, and could most probably be further improved through the optimization of the concentration of charge carriers by doping. The mechanism suggested in study opens new paths to improve the thermoelectric performances of other families of materials.

## 4. Experimental Section

$AgSnSe_2$ was synthesized by melting and solidifying the raw materials Ag (shots, 99.99%), Sn (shots, 99.99%) and Se (granules, 99.99%) together at 1073 K in argon-flushed and evacuated silica tubes. The ingots were dwelled at 1073 K for 12 h, cooled down to 673 K in 8 h, and then water-quenched. The samples were then ground by hand and pressed into pellets under uniaxial stress (250 MPa), and the resulting pellets were heated again with the same thermal treatment. PbSe was synthesized by melting and solidifying the raw materials Pb (granules, 99.99%) and Se (granules, 99.99%) together at 1423 K for 6 h in argon-flushed and evacuated silica tubes. $AgPb_mSnSe_{m+2}$ were synthesized by mixing, melting and solidifying previously prepared $AgSnSe_2$ and PbSe together with mole ratio 1: m at 1423 K in argon-flushed and evacuated silica tubes. The ingots were dwelled at 1423 K for 6 h, cooled down to 673 K in 8 h, and then water-quenched. The obtained samples were ground and subsequently densified by using a spark plasma sintering (SPS) system (SPS-511S) at 873 K with holding time of 10 min in a $\varnothing$=15 mm graphite mold under an uniaxial compressive stress of 100 MPa in an argon atmosphere. The resulting graphite contamination of the surface was removed by polishing, and the samples were then cut to the dimensions appropriate for the measurements using a diamond saw. For thermal stability tests, the pellet with composition $AgPb_mSnSe_{2+m}$ (m = 25) used for electrical properties measurements was annealed at 773 K for one week in an evacuated silica tube before being measured again.

Room temperature X-ray diffraction characterization was performed using a Panalytical X'Pert diffractometer by using a Cu Kα1 radiation obtained using a Ge(111) incident monochromator and an X'celerator detector. Rietveld refinement was performed using FULLPROF software[27]. Scanning electron microscopy (SEM) studies were performed using JSM-6510, and energy dispersive spectrometer (EDS) was performed using ThermoFisher NS7 EDS.

The specific heat $C_p$ was measured using a Quantum Design PPMS (physical properties measurement system) from 400 K to 2 K, with the thermal contact ensured by using apiezon H thermal grease between 400 K and 290 K and apiezon N between 300 K and 2 K (the agreement between both measurement series was very good as it can be observed in figure 4a). Hall effect measurements were performed in the PPMS environment from 400 K to 20 K using a Keithley 6220 current source and a Keithley 2182A nanovoltmeter with square samples. The hall coefficient was obtained from the linear fit of the Hall resistivity between -9 T and 9 T. The low temperature Seebeck coefficient $S$ and the electrical resistivity $\rho$ were measured simultaneously by differential method with two T-type thermocouples by using the slope of ΔV-ΔT curve with gradients up to about 0.2 K/mm, by using a laboratory made system in a He-free cryostat from 20 K to room temperature (R.T.). From 340 K to 889 K, the Seebeck coefficient $S$ and the electrical resistivity $\rho$ were measured simultaneously using a commercial system (LSR-3, Linseis) in helium atmosphere. All electrical characterizations were performed on bars cut from the pellets in a direction perpendicular to the pressing direction. Thermal conductivity values from 300 K to 873 K were calculated from $\lambda = C_p\, k\, d$, where thermal diffusivity coefficient $k$ in a direction parallel to the pressing direction of (8 mm x 8 mm) square pellets was obtained using a laser flash method (Netzsch LFA 427). $C_p$ values were almost temperature and composition independent above 200 K and lay within a few percent of the calculated Dulong-Petit value. As a very slight increase of $C_p$ was observed between 200 K and 400 K, a linear extrapolation was used to calculate $\lambda$ and ZT (corresponding to a 10% increase of $C_p$ beween 300 K and 873 K), which most probably lead to slightly pessimistic values. The combined uncertainty for all measurements involved in the calculation of *ZT* is about 15%.

**Supporting Information**
Supporting Information is available from the Wiley Online Library or from the author.

Rietveld refinements of AgPb$_m$SnSe$_{2+m}$ (m = ∞, 100, 50, 25) at room temperature (Figure S1), Low magnification SEM and EDX of the fractured surfaces of typical sample AgPb$_m$SnSe$_{2+m}$ (m = 50) (Figure S2), Temperature dependences of (a) electrical resistivity and (b) Seebeck coefficient of AgPb$_m$SnSe$_{2+m}$ (m = 25) from 345 K to 889 K (Figure S3).

**Acknowledgements**
This work was supported by the Natural Science Foundation of China under Grant No.51272103. This work was also supported by the "Zhuoyue" Program from Beihang University, the Recruitment Program for Young Professionals and the Natural Science Foundation of China under Grant No.51571007.

Received: ((will be filled in by the editorial staff))
Revised: ((will be filled in by the editorial staff))
Published online: ((will be filled in by the editorial staff))

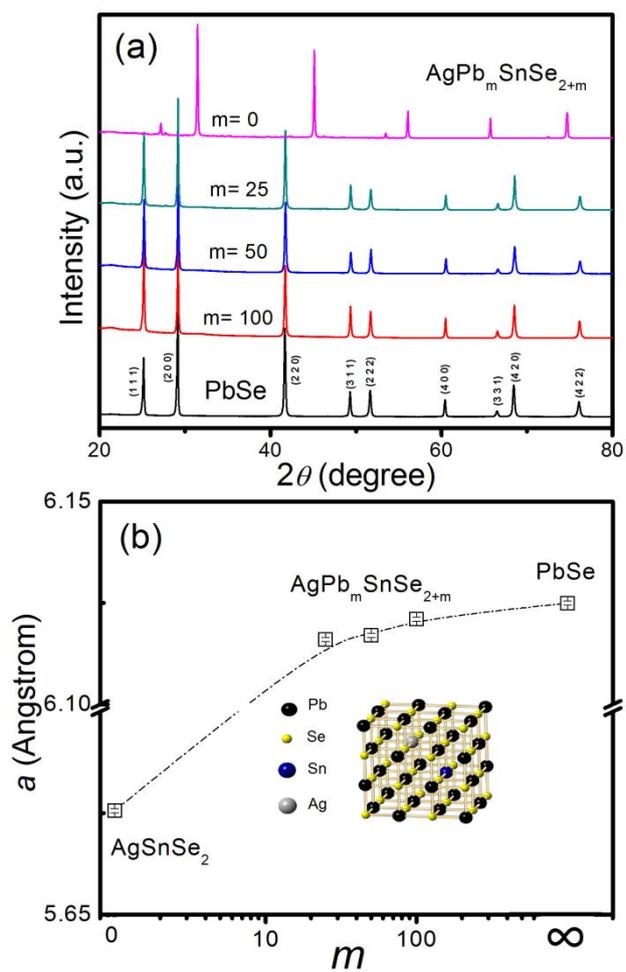

**Figure 1.** Powder XRD patterns (a) and lattice constants (b) of AgPb$_m$SnSe$_{2+m}$ (m = ∞, 100, 50, 25, 0).

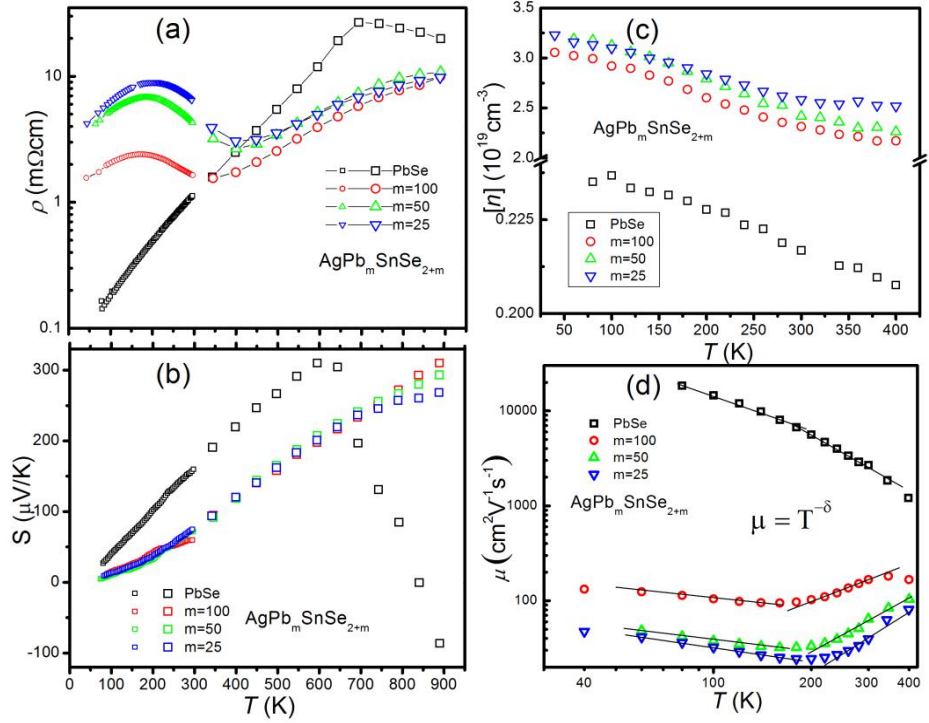

**Figure 2.** Temperature dependences of (a) electrical resistivity, (b) Seebeck coefficient, (c) carriers concentration and (d) hole mobility of AgPb$_m$SnSe$_{2+m}$ (m = ∞, 100, 50, 25).

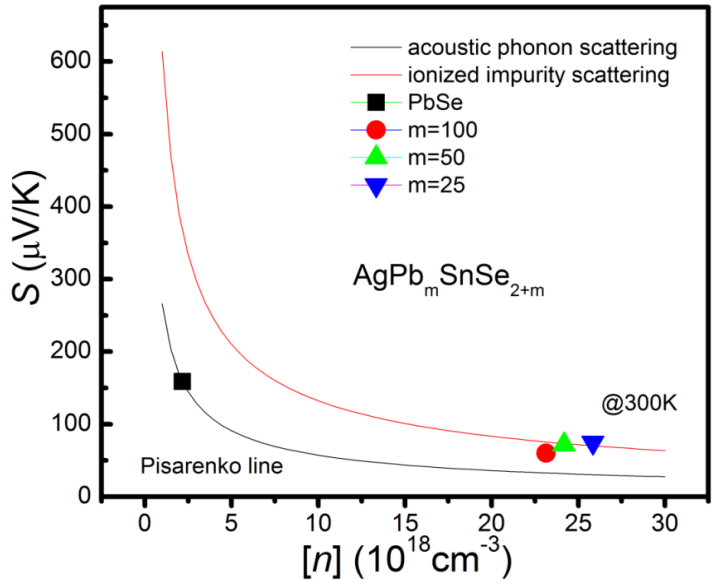

**Figure 3.** Room-temperature Pisarenko plot for AgPb$_m$SnSe$_{2+m}$ (m = ∞, 100, 50, 25). The solid black and red lines guide for eyes in this figure are based on a model employing a single parabolic band with an effective mass $m^* = 1.3\ m_e$ with acoustic phonon scattering mechanism ($r = -1/2$), and $m^* = 1\ m_e$ with ionized impurity scattering mechanism ($r = 3/2$), respectively.

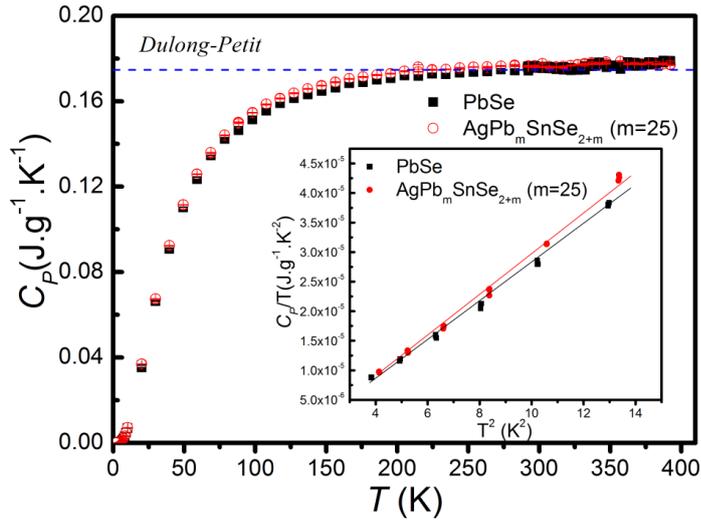

**Figure 4.** The heat capacity $C_p$ of PbSe and AgPb$_m$SnSe$_{2+m}$ (m = 25) as a function of temperature from 2 K to 400 K. The inset shows the low-T $C_P$ of PbSe and AgPb$_m$SnSe$_{2+m}$ (m = 25) plotted as $C_p$ / T versus T$^2$.

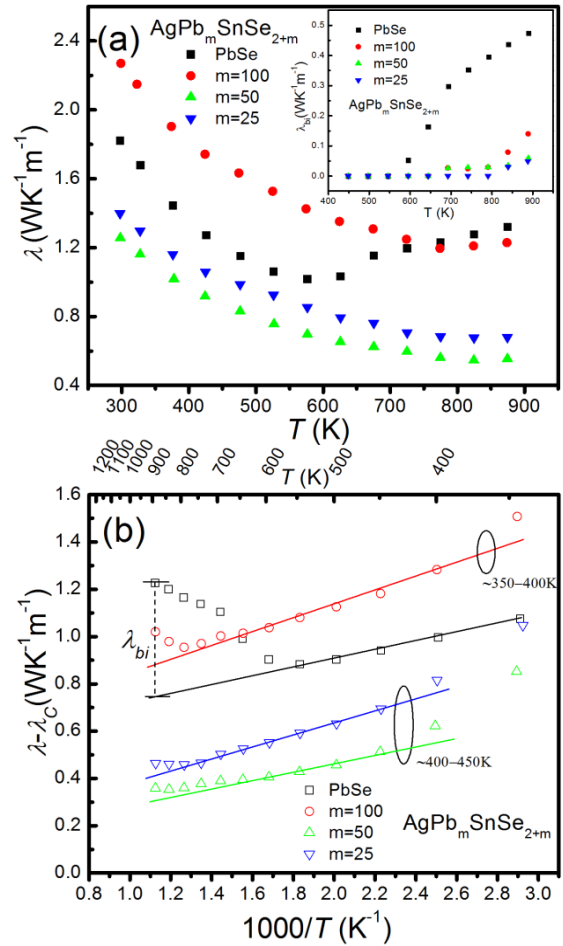

**Figure 5.** (a) Temperature dependence of thermal conductivity of AgPb$_m$SnSe$_{2+m}$ (m = ∞, 100, 50, 25) from 300 K to 875 K. The inset in (a) shows bipolar thermal conductivity as a function of temperature for AgPb$_m$SnSe$_{2+m}$ (m = ∞, 100, 50, 25). (b) The difference of total and carrier thermal conductivity $\lambda - L\sigma T$ as a function of temperature for AgPb$_m$SnSe$_{2+m}$ (m = ∞, 100, 50, 25), the solid line is a linear fit of the lattice thermal conductivity at temperature ranges from room temperature to 875 K, deviation of thermal conductivity indicating a significant bipolar thermal conductivity.

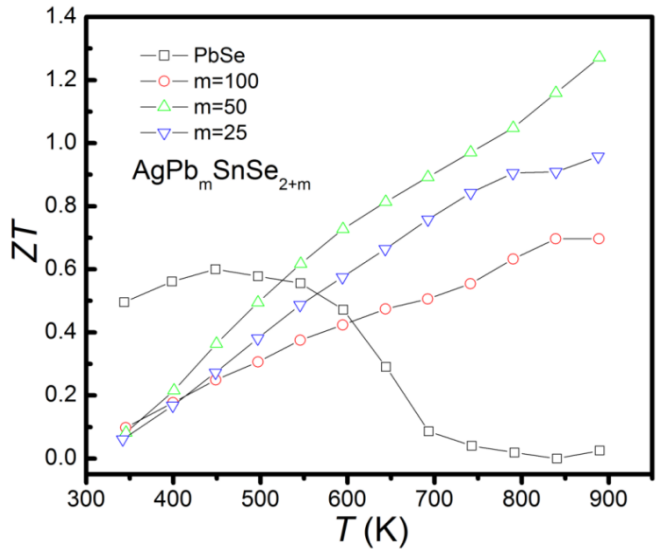

**Figure 6.** ZT of AgPb$_m$SnSe$_{2+m}$ (m = ∞, 100, 50, 25) in the range 340 K to 890 K.

**Table 1.** carrier concentration [$n$] and holes mobility $\mu$, the power exponent $\delta$ of $\mu \approx T^\delta$, the main scattering mechanism $r$ (acoustic phonon scattering and ionized impurity scattering abbreviate as AP and II, respectively), transport effective mass $m^*(m_e)$, and relaxation time $\tau$ for AgPb$_m$SnSe$_{2+m}$ (m = 100, 50, 25) and pristine PbSe at room temperature.

| AgPb$_m$SnSe$_{2+m}$ | [$n$] ($10^{19}$ cm$^{-3}$) | $\mu$ (cm$^2$V$^{-1}$s$^{-1}$) | $\delta$ | $r$ | $m^*(m_e)$ | $\tau(10^{-14}$ S) |
|---|---|---|---|---|---|---|

| PbSe | 0.22 | 2690 | 1.8 | AP | 1.3 | 199 |
| --- | --- | --- | --- | --- | --- | --- |
| $m = 100$ | 2.31 | 167 | -1.4 | II | 0.79 | 7.5 |
| $m = 50$ | 2.42 | 64 | -1.5 | II | 0.98 | 3.6 |
| $m = 25$ | 2.58 | 40 | -1.7 | II | 1.06 | 2.4 |

Supporting Information

**The role of ionized impurity scattering on the thermoelectric performances of rock salt AgPb$_m$SnSe$_{2+m}$**

*Lin Pan$^{1, 4*}$, Sunanda Mitra$^2$, Li-Dong Zhao$^3$, Yawei Shen$^1$, Yifeng Wang$^{1*}$ Claudia Felser$^4$ and David Berardan$^{2*}$*

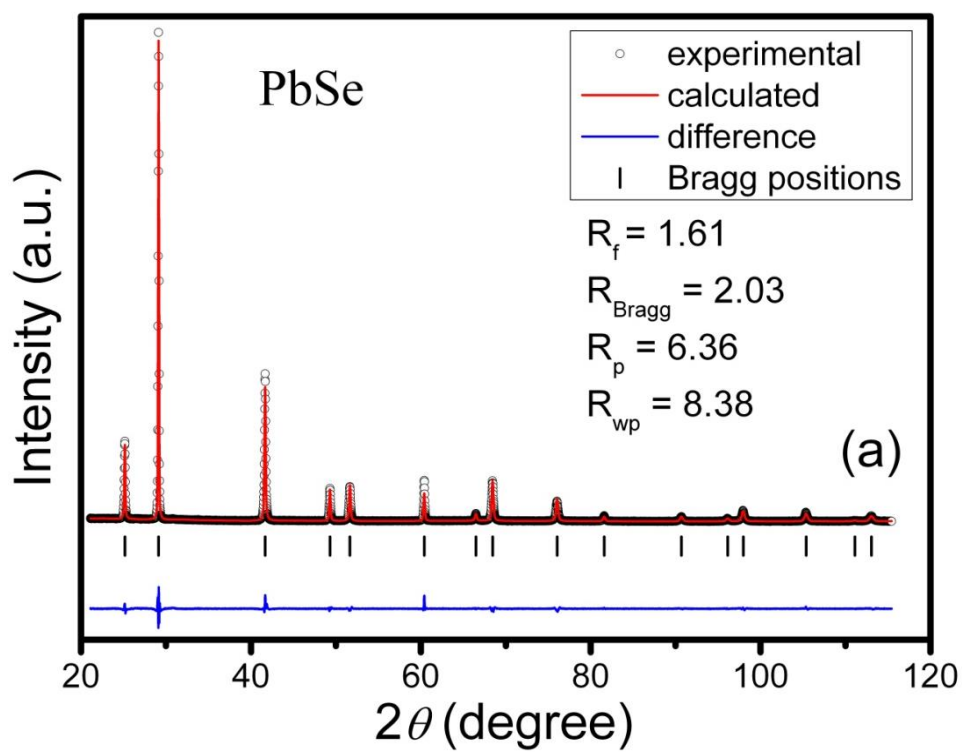

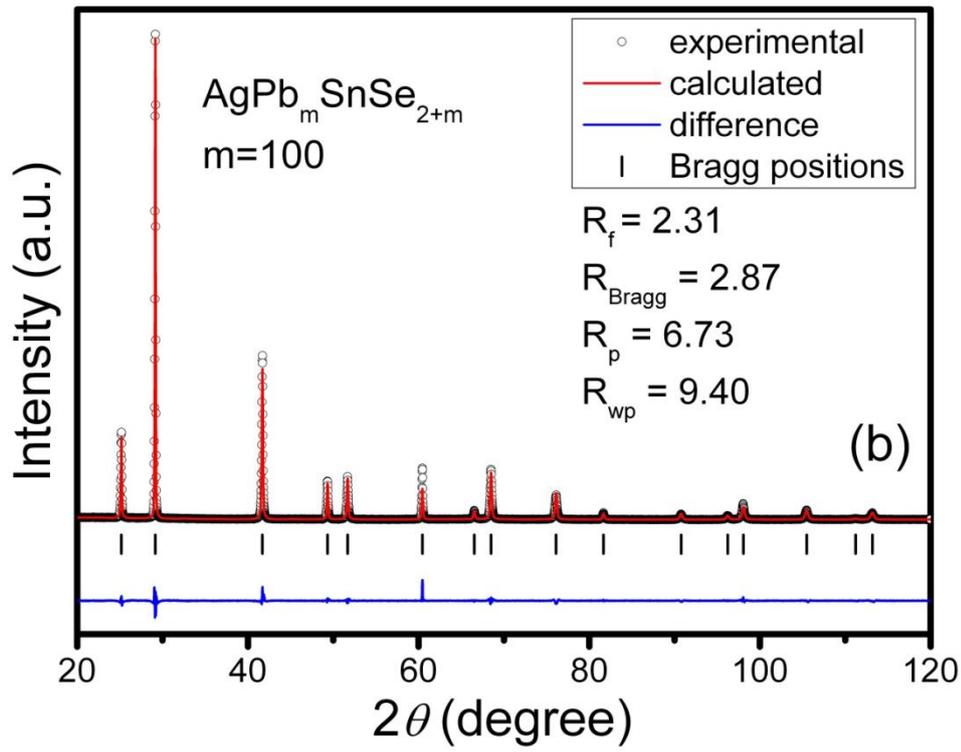

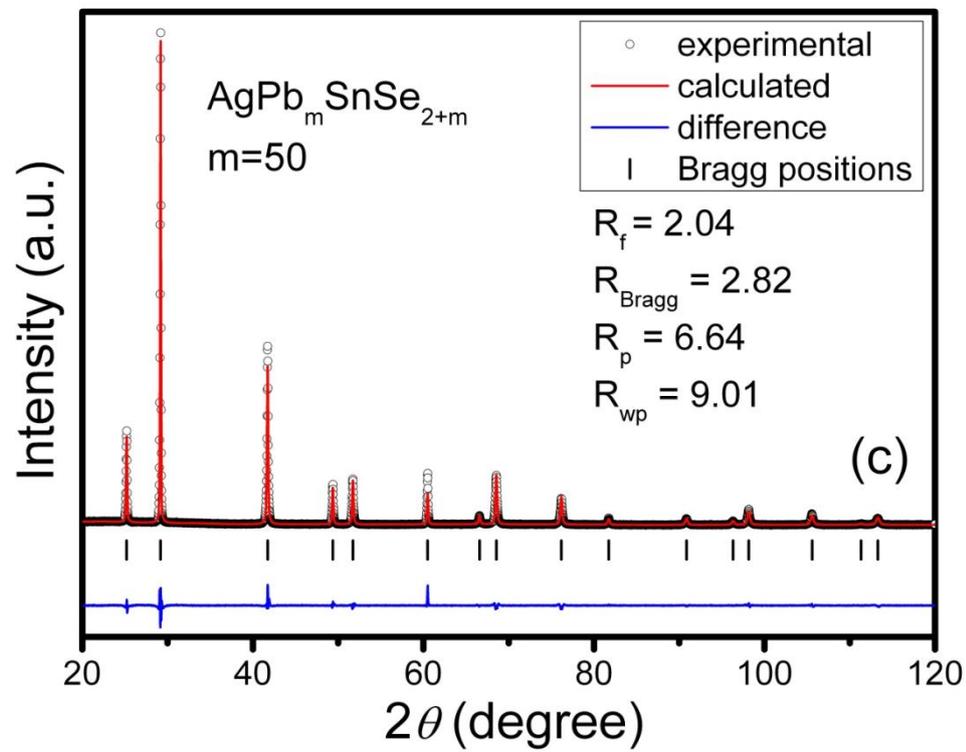

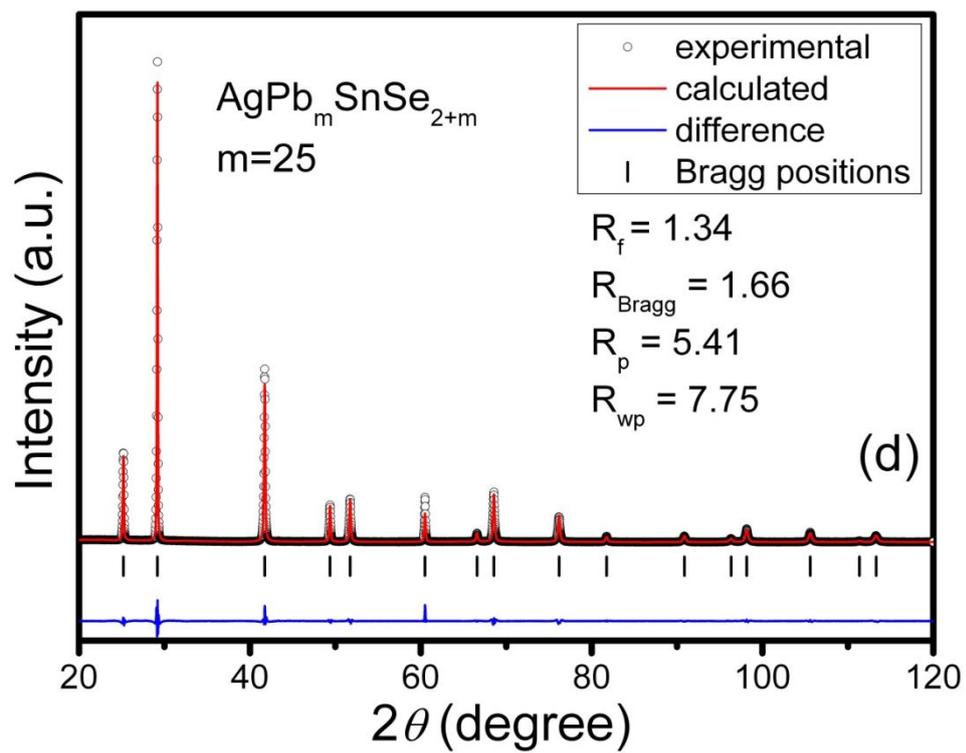

Figure S1 Rietveld refinements of AgPb$_m$SnSe$_{2+m}$ (m = ∞, 100, 50, 25) at room temperature.

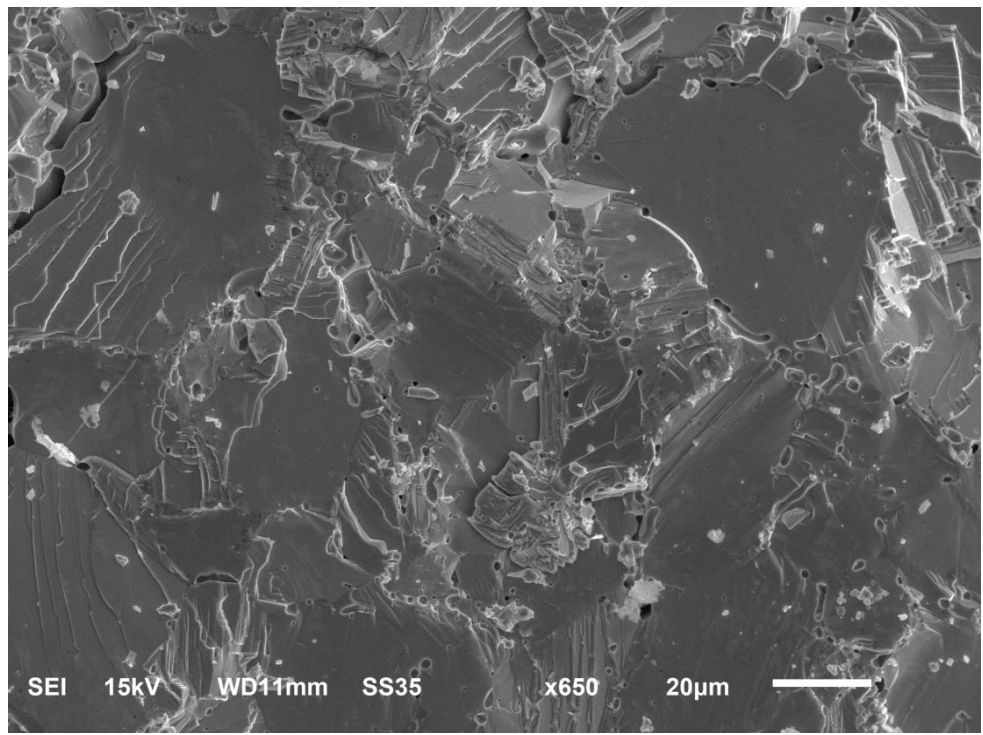

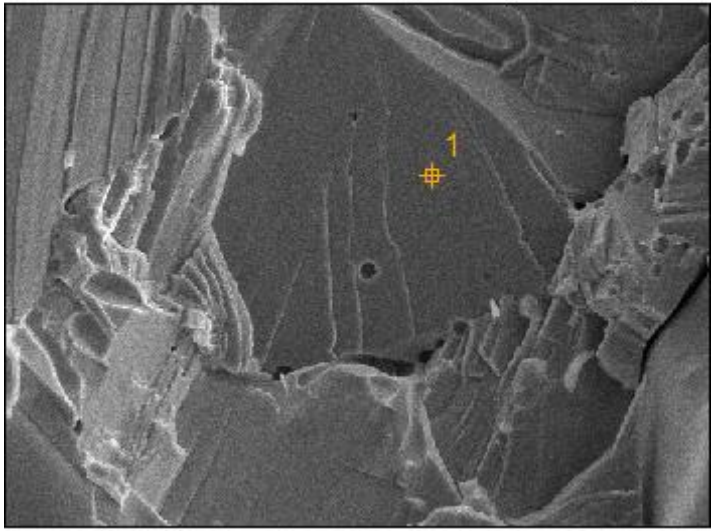

```
Image Name:       01-03
Image Resolution: 512 by 384
Image Pixel Size: 0.13 µm
Acc. Voltage:     15.0 kV
Magnification:    1999
```

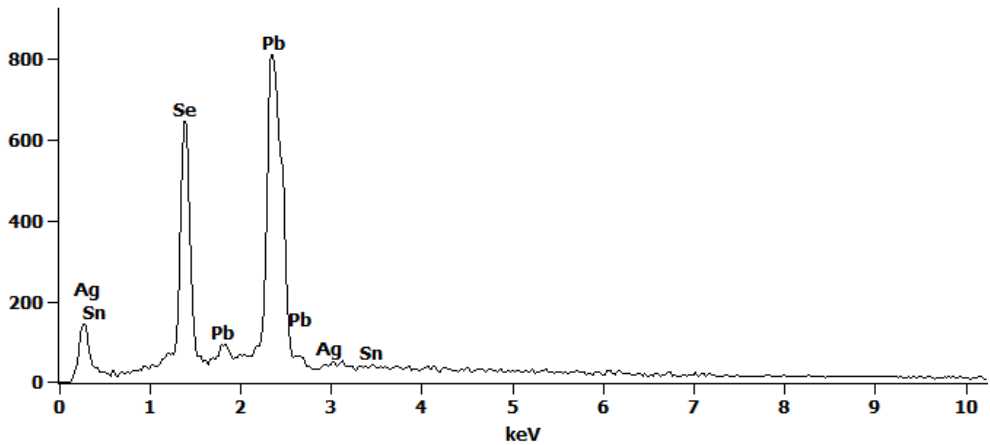

Weight %

|  | Se-L | Ag-L | Sn-L | Pb-M |
|---|---|---|---|---|
| 01-03_pt1 | 26.27 | 1.01 | 0.61 | 72.11 |

Atom %

|  | Se-L | Ag-L | Sn-L | Pb-M |
|---|---|---|---|---|
| 01-03_pt1 | 47.86 | 1.35 | 0.73 | 50.06 |

Formula

|  | Se-L | Ag-L | Sn-L | Pb-M |
|---|---|---|---|---|
| 01-03_pt1 | Se | Ag | Sn | Pb |

Compound %

| | Se | Ag | Sn | Pb |
|---|---|---|---|---|
| 01-03_pt1 | 26.27 | 1.01 | 0.61 | 72.11 |

Figure S2 Low magnification SEM and EDX of the fractured surfaces of typical sample

$AgPb_mSnSe_{2+m}$ (m = 50)

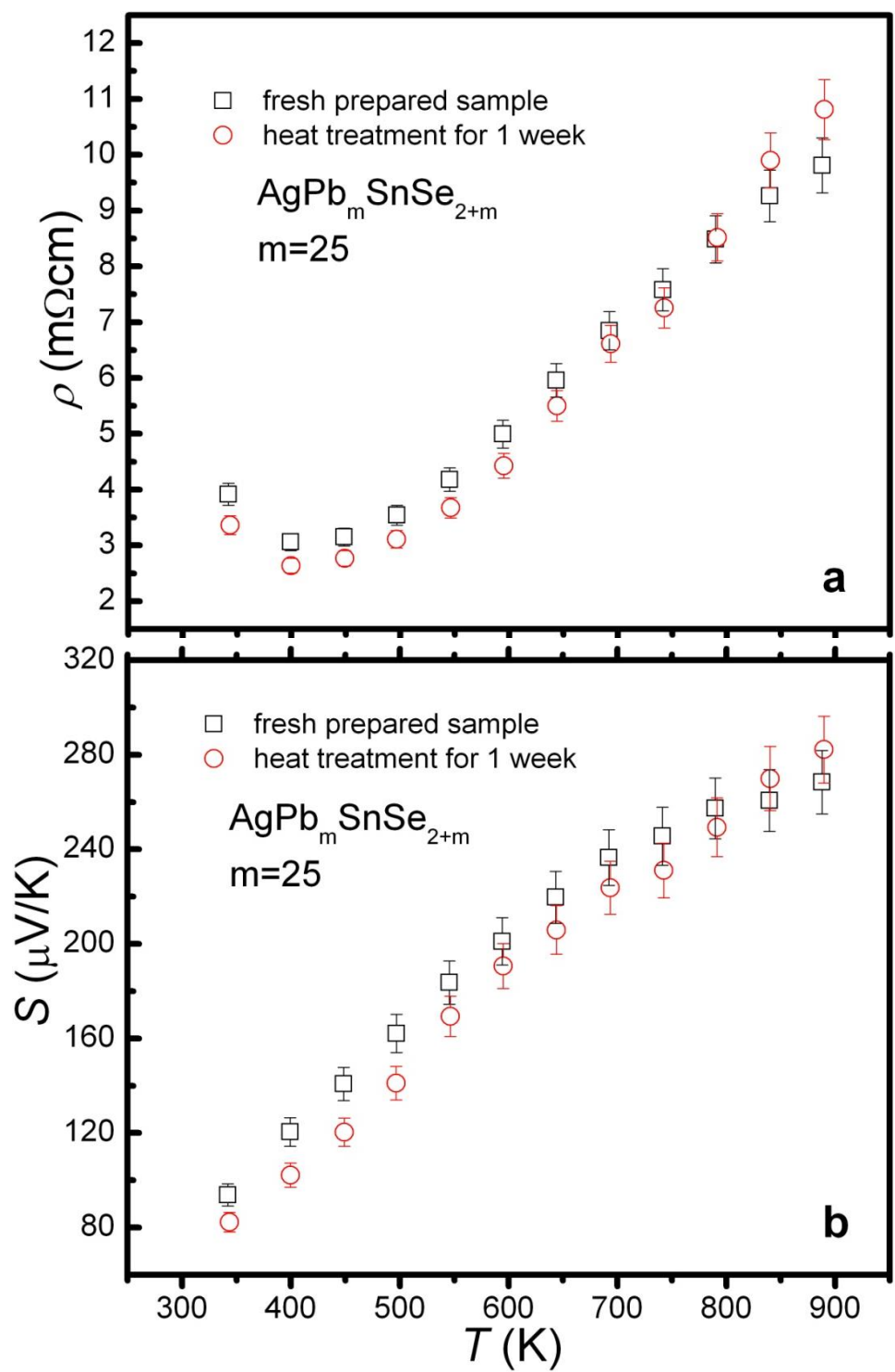

Figure S3 Temperature dependences of (a) electrical resistivity and (b) Seebeck coefficient of

$AgPb_mSnSe_{2+m}$ (m = 25) from 345 K to 889 K.